\documentclass[12pt]{article}
\usepackage{times}
\usepackage{geometry}
\usepackage[utf8]{inputenc}
\usepackage{enumitem,amssymb}
\usepackage{ragged2e}
\usepackage{graphicx,xspace,lscape,floatflt,wasysym}
\usepackage{graphicx,wasysym}
\usepackage{natbib}
\usepackage{multicol}
\usepackage{microtype} 

\newlist{thematic}{itemize}{8}
\setlist[thematic]{label=$\square$}
\usepackage{pifont}

\begin{document}
\raggedright
\huge
Astro2020 Science White Paper \linebreak

Astromineralogy of interstellar dust with X-ray spectroscopy \linebreak
\normalsize

\noindent \textbf{Thematic Areas:} \hspace*{60pt} $\square$ Planetary Systems \hspace*{10pt} $\XBox$ Star and Planet Formation \hspace*{20pt}\linebreak
$\square$ Formation and Evolution of Compact Objects \hspace*{31pt} $\square$ Cosmology and Fundamental Physics \linebreak
  $\square$  Stars and Stellar Evolution \hspace*{1pt} $\square$ Resolved Stellar Populations and their Environments \hspace*{40pt} \linebreak
  $\square$    Galaxy Evolution   \hspace*{45pt} $\square$             Multi-Messenger Astronomy and Astrophysics \hspace*{65pt} \linebreak
  
\textbf{Principal Author:}

Name: L\'{i}a Corrales	
 \linebreak						
Institution: University of Michigan 
 \linebreak
Email: liac@umich.edu
 \linebreak
Phone: (734) 763-8915 
 \linebreak
 
\textbf{Co-authors:} 
Lynne Valencic (Johns Hopkins University); 
Elisa Costantini (SRON); 
Javier Garc\'{i}a (Caltech); 
Efrain Gatuzz (MPA); 
Tim Kallman (GSFC); 
Julia Lee (Harvard); 
Norbert Schulz (MIT); 
Sascha Zeegers (ASIAA); 
Claude Canizares (MIT);  
Bruce Draine (Princeton); 
Sebastian Heinz (University of Wisconsin); 
Edmund Hodges-Kluck (GSFC); 
Edward B.~Jenkins (Princeton); 
Frits Paerels (Columbia);  
Randall K.~Smith (SAO); 
Tea Temim (STScI); 
J\"{o}rn Wilms (University of Erlangen-Nuremberg); 
Daniel Wolf~Savin (Columbia)
  \linebreak

\justifying
\textbf{Abstract  (optional):}

X-ray absorption fine structure (XAFS) in the 0.2--2~keV band is a crucial component in  multi-wavelength studies of dust mineralogy, size, and shape -- parameters that are necessary for interpreting astronomical observations and building physical models across all fields, from cosmology to exoplanets. 
Despite its importance, many fundamental questions about dust remain open. 
What is the origin of the dust that suffuses the interstellar medium (ISM)?  
Where is the missing interstellar oxygen? How does iron, predominantly produced by Type~Ia supernovae, become incorporated into dust? 
What is the main form of carbon in the ISM, and how does it differ from carbon in stellar winds? 
The next generation of X-ray observatories, employing microcalorimeter technology and $R \equiv \lambda/\Delta \lambda \geq 3000$ gratings, will provide pivotal insights for these questions by measuring XAFS in absorption and scattering. 
However, lab measurements of mineralogical candidates for astrophysical dust, with $R > 1000$, are needed to fully take advantage of the coming observations. 

\pagebreak

The effects of cosmic dust can be seen in virtually every field of astrophysics. 
The cycle of baryons, the growth of molecules, the physics of cosmic gas, the formation of planets, and the origin of life all rely upon the formation of solid phase materials in space. 
Cosmic dust has even been found in extraordinarily unexpected places: from the hot halo gas surrounding galaxies \citep{Engelbracht2006, Menard2010} to the filaments of molecular gas in galaxy clusters \citep{RussellH2014, RussellH2017b}.

Despite the importance, ubiquity, and decades of study on astrophysical dust, shockingly basic questions remain. 
From the perspective of timescales, even the presence of dust in the diffuse interstellar medium (ISM) remains a mystery. All dust grains injected by AGB stars into the surrounding ISM 
will eventually be 
processed by the destructive force of a supernova shock \citep{Dwek2008,Jones2013, Raymond2013}.  However, the majority of refractory elements that become incorporated into interstellar dust -- Mg, Si, and Fe -- are produced in supernovae explosions \citep{Dwek2016}. 
The dust that formed rapidly in the ejecta may be destroyed later by the reverse shock, which takes tens of thousands of years to fully heat the remnant \citep{Gall2011, Raymond2013, Gall2014}.
The debate continues whether supernovae – on average – produce or destroy dust.  
The answer will determine what fraction of  interstellar solid matter is formed in the low density environment of the diffuse ISM, through deposition on existing grain surfaces \citep{Draine2009}.
\textbf{Thus, the origin and fate of dust in the Universe remains a fundamental question in astrophysics, which can be answered by measuring the composition, size, and structure of interstellar dust.}

High resolution X-ray spectroscopy provides abundances for all ions, from
neutral to H-like species, in gaseous and solid form, over a broad range of ISM densities
($N_{\rm{H}} \sim 10^{20}$--$10^{24}\,\mathrm{cm}^{-2}$).
The signature of solid phase minerals and glasses are imprinted in the shape of the absorption features in high resolution X-ray spectra \citep{LR2005,Lee2009}. Studying these X-ray absorption fine structure (XAFS) 
features yield the constituent minerals, crystalline content (versus amorphous), size, and shape of interstellar dust grains. 
\textbf{X-ray spectroscopy can provide a complete inventory of both the gas and solid phase of the ISM, while directly identifying the mineral building-blocks of cosmic dust.}

\vspace{-2.5ex}
\section{Open questions in astromineralogy}
\label{sec:astromineralogy}
\vspace{-1ex}

Astromineralogy rests upon two major pillars of observational astronomy. 
First, direct gas phase abundances found via UV/optical absorption line studies, compared to a total abundance table, determine the fraction in the solid phase \citep{Jenkins2009}. These studies are limited to low opacity sight lines as UV/optical light cannot penetrate the dense ISM. 
Second, IR and radio studies detect polycyclic aromatic hydrocarbons (PAHs), silicates, and ices in the dense ISM. These species are studied using rotational and vibrational transitions, which require complex models.
Determining the abundances or grain structures based on these features  requires making strong assumptions about their physical properties \citep{Draine2003}.

High throughput, high resolution X-ray spectroscopy can reveal both the structure and composition of interstellar dust. 
Structures in X-ray spectra imposed by XAFS provide a direct measure of the spacing between atoms in a crystalline
lattice \citep{LR2005}, while lab measurements of astrophysically relevant materials provide templates for measuring dust components \citep{Lee2009, Zeegers2017, Rogantini2018}. 
\textbf{To fully interpret near edge X-ray photoabsorption features, optical constants derived from lab absorption measurements must be incorporated into models that calculate the grain size- and shape-dependent scattering properties} \citep{HD2016, Corrales2016, Zeegers2017}. 

Progress in studying the cold phases of the ISM through spectroscopic capabilities of \textsl{Chandra} and \textsl{XMM-Newton} ($R \equiv \lambda / \Delta \lambda \sim 500-1000$ with effective areas $A_{e} \sim 100\,\mathrm{cm}^2$), made over the last two decades, has yielded results that are inconsistent with the standard picture of ISM dust as studied at longer wavelengths. We review open problems and X-ray findings that require new observations, lab measurements, and theoretical calculations to solve.

\begin{figure}[htp]
    \centering
    \includegraphics[width=0.49\textwidth]{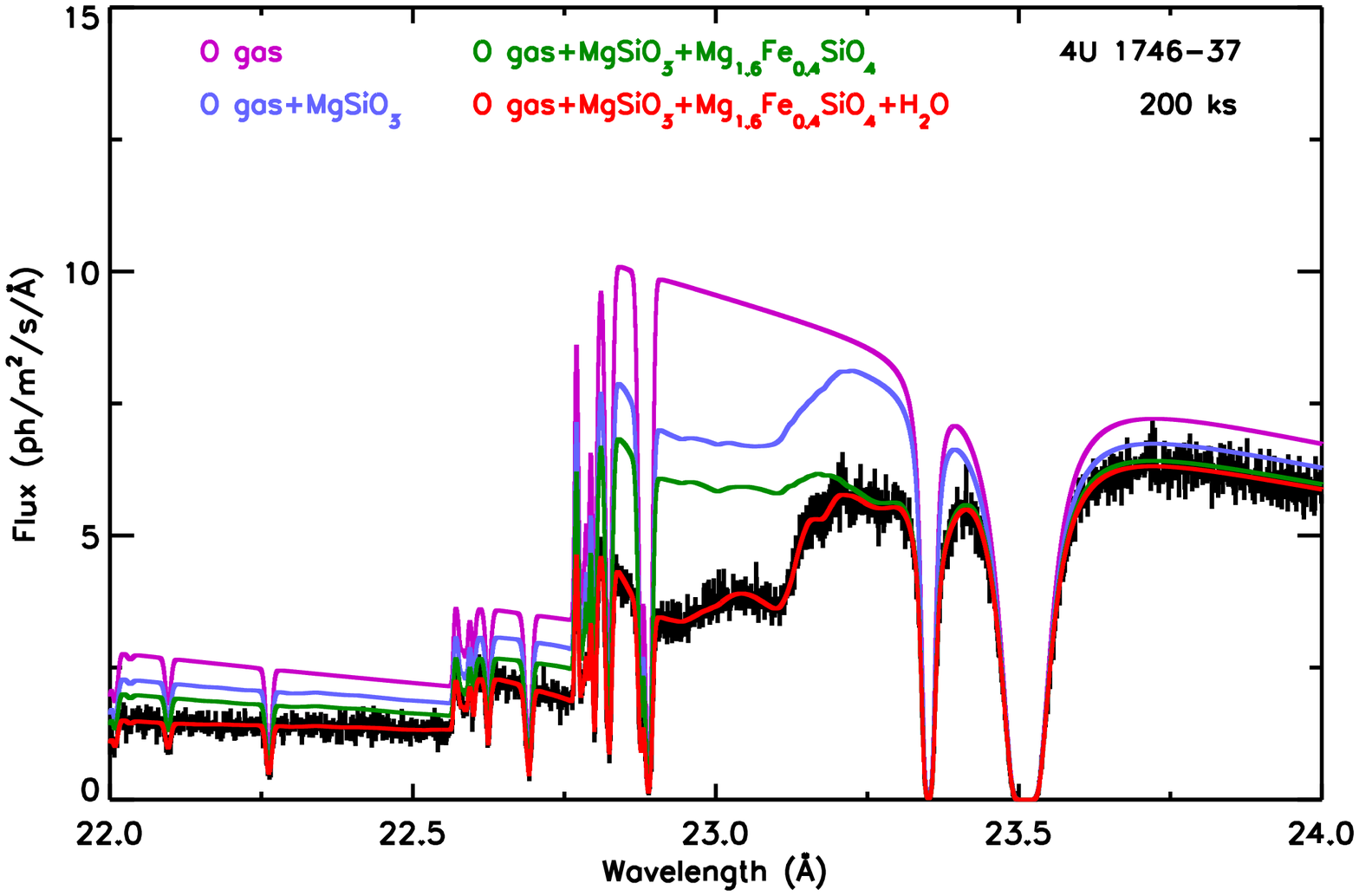}
    \includegraphics[width=0.49\textwidth]{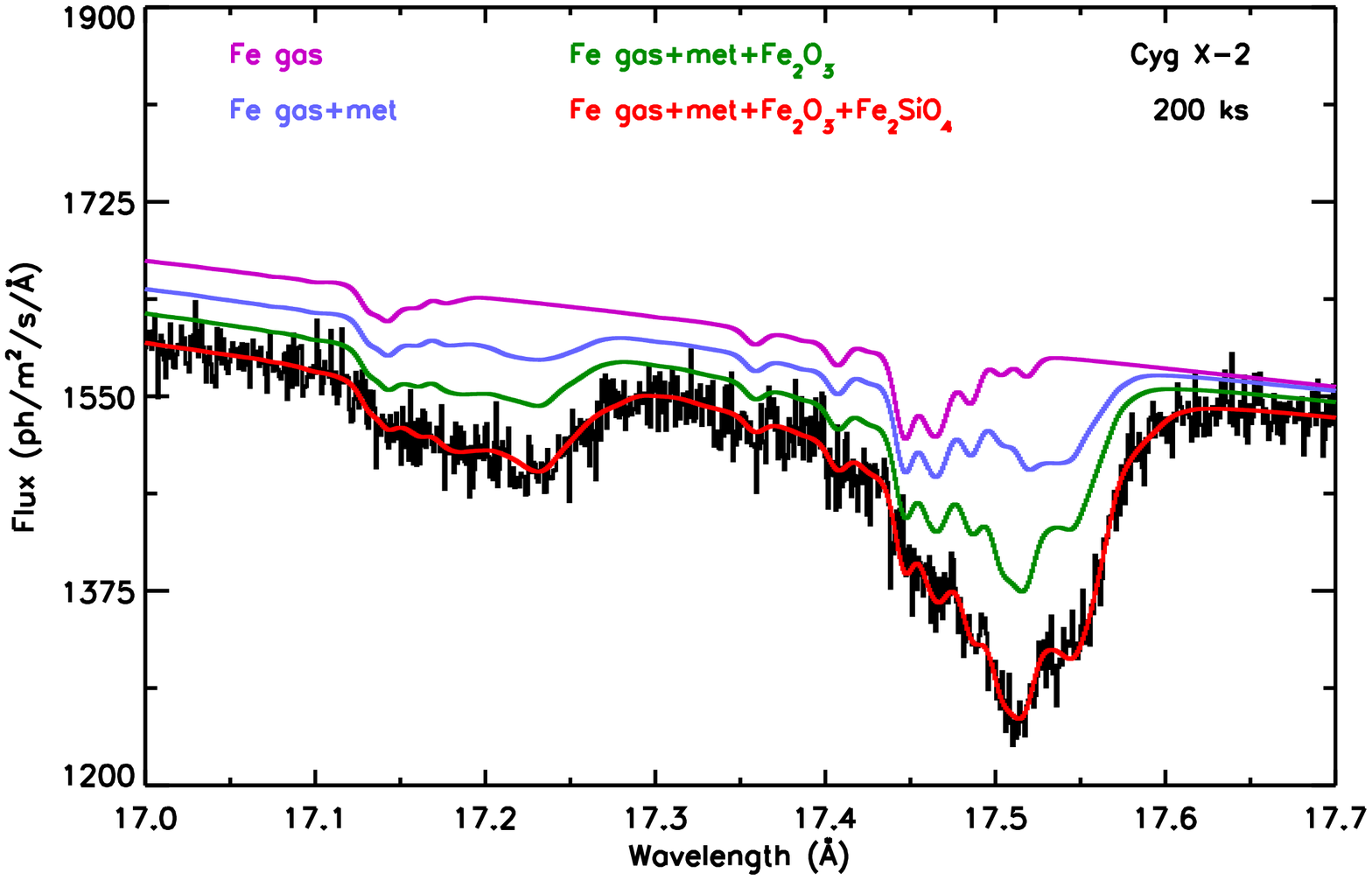}
\caption{
Simulated spectra showing the effects of absorption due to dust.
\textsl{Left:} 4U 1746-37 ($N_\mathrm{H}=5\times10^{21}\,\mathrm{cm}^{2}$) at the oxygen K edge as observed with a telescope with $R=5000$ and $A_{e}=4000\,\mathrm{cm}^{2}$. If water ice on grains near the diffuse ISM/cloud interface accounts for the ``missing oxygen'', this mission will detect it. 
\textsl{Right:} Cyg X-2 ($N_\mathrm{H}=10^{22}\,\mathrm{cm}^2$) at the Fe L edge, observed with $R=3000$ and $A_{e}=500\,\mathrm{cm}^2$. It will easily distinguish between iron in various species of magnetic inclusions or silicates, important for grain polarization at longer wavelengths.
}
    \label{fig:OandFespectrum}
\end{figure}

\vspace{-3ex}
\subsection{Solving the Oxygen problem (0.5\,keV)}
\label{sec:Oxygen}
\vspace{-1ex}

Oxygen is the most abundant interstellar metal and is a key ingredient of ices and silicate minerals. Studies of the gas-phase ISM suggest that ${\sim}30\%$ of oxygen is locked up in solids, a fraction that increases with decreasing ISM temperatures. In the coldest regions, more oxygen is ``missing'' from the gas phase than can be explained by silicate minerals alone, suggesting the presence of icy grain mantles \citep{Jenkins2009}. \citet{Poteet2015} have shown that the H2O ice would have to be in very thick mantles to have escaped detection by IR spectroscopy. Oxygen is thereby an ideal target for X-ray absorption studies on sightlines where there is missing oxygen. Studies will need to take into account the effects of large grain sizes on the X-ray extinction cross sections, but these can be modeled. 

Early measurements of oxygen X-ray absorption argued for the appearance of additional K shell absorption edges \citep{Paerels2001, Takei2002} and residual features arising from the solid phase \citep{deVries2009}. Later studies demonstrated that many of these features could be explained by absorption from gas-phase O\textsc{ii} and O\textsc{iii}, and by including the effects of Auger decay \citep{Juett2004, Garcia2011, Gorczyca2013, Gatuzz2015}. A recent study by \citet{Joachimi2016} also failed to find statistically relevant CO features in X-ray absorption from a majority of X-ray binary sight lines. 
{\bf Resolving this disagreement between X-ray observations and the canon of ISM dust studies at longer wavelengths is crucial for deciphering the mystery of ``missing oxygen.''
}

\begin{floatingfigure}[r]{3.3in}
\includegraphics[totalheight=2.0in]{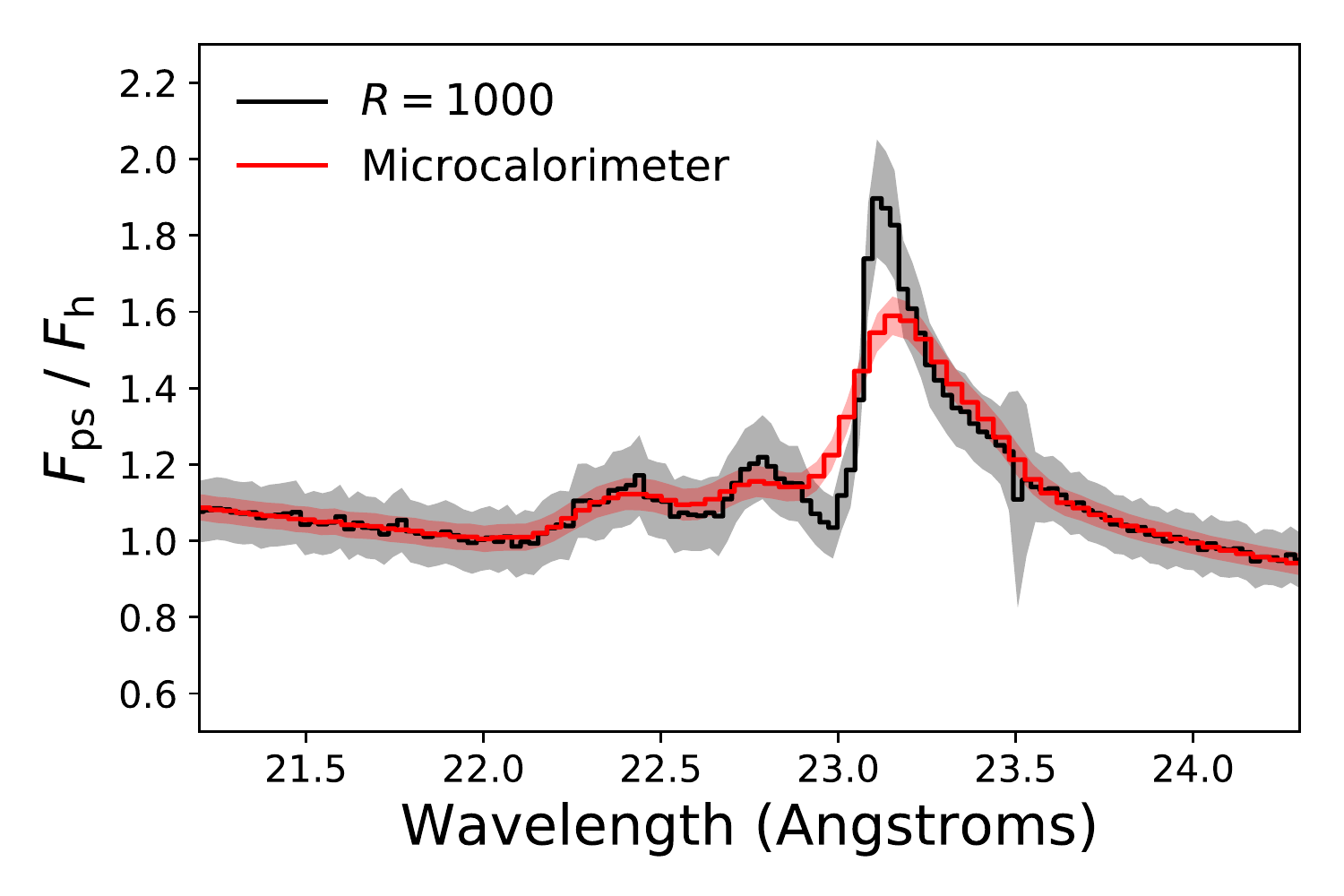}
\vspace{-0.3in}
\caption{Example 50~ks astrosilicate O K \citep{Draine2003} measurement obtained from the ratio of a point source ($F_{ps}$) and halo spectrum ($F_h$) using $A_{e} = 500$~cm$^2$, $R = 1000$ (black) or a microcalorimeter ($R \approx 250$, $A_{e}= 10^4$~cm$^2$, in red).}
\label{fig:OKhalo}
\end{floatingfigure}

The currently available cross-sections with XAFS for O\textsc{i} K-shell absorption are incorporated in the \texttt{amol} model in SPEX \citep{SPEXcode}.
These cross-sections are available at a variety of resolutions, from $R \leq 1000$ \citep[O$_2$, CO, CO$_2$, and heavy metal oxides, ][]{Barrus1979, vanAken1998} to $R \geq 5000$ \citep[H$_2$O, ][]{Hiraya2001, Parent2002}. 
Figure~\ref{fig:OandFespectrum} (left) shows the O K-edge from a flagship mission that would be able to measure the abundance of various oxygen bearing species along highly absorbed sight lines in 200\,ks.

X-ray missions using microcalorimeter technology, creating the X-ray observational equivalent of an integral field spectrograph, will enable high resolution spectroscopy of dust scattering halos, which will exhibit resonant structure around the photoelectric edge (Figure~\ref{fig:OKhalo}, see also white paper by Valencic et al., 2019). 
However, these features cannot be fully interpreted 
until high resolution lab measurements of XAFS from ices (containing CO, CO$_2$, CH$_3$OH, and so on) at $R \geq 1000$, and the corresponding optical constants, are obtained. 

\vspace{-2.5ex}
\subsection{How is iron incorporated into the solid phase? (0.7~keV)}
\label{sec:Iron}

\vspace{-1ex}

The mystery of how iron is incorporated into dust -- its dominant form -- has vexed astronomers for decades. 
Type Ia supernovae produce the majority of interstellar iron but do not appear to produce dust. Therefore, iron dust must form rapidly in the ISM \citep{Dwek2016}. 
Olivine is a prime candidate material for interstellar silicates, and can exhibit a range of Mg to Fe ratios: fayalite contains iron only (Fe$_2$SiO$_4$) while forsterite contains magnesium only (Mg$_2$SiO$_4$). 
Learning whether interstellar iron is bound mostly in silicates, oxides, or magnetic nanoparticle inclusions will uncover crucial dust formation mechanisms \citep[e.g., simulations by][]{Zhukovska2016,Zhukovska2018}.

This question has taken on even greater importance recently, as attempts to detect the cosmic microwave background polarization have  shown how little we know about the polarized dusty Galactic foreground \citep{Barkats2014,Remazeilles2016}. 
The polarized emission from foreground dust can change quickly in the microwave regime and is highly dependent on grain size and ferric composition, whether iron is primarily in magnetic inclusions (e.g., Fe$_{2}$O$_{3}$, Fe$_{3}$O$_{4}$) or silicates \citep{Draine2009b, Draine2013, Hoang2016}. 

A MIDEX-class detector with spectral resolution $R=3000$ and $A_{e} = 500~\mathrm{cm}^{2}$ will be able to distinguish between different forms of iron, 
as seen in the Fe L edge region simulated with the \texttt{amol} model from SPEX (Figure~\ref{fig:OandFespectrum}, right).
The current lab measurements of the Fe L region are available with $R \approx 3500$ \citep{Lee2009}, but many commonly used X-ray absorption models (\texttt{tbvarabs}, \texttt{ISMabs}) incorporate the Fe L edge of pure solid iron at $R \sim 500$ \citep{KK2000}.   
New lab measurements of Fe~K shell (7.1\,keV) absorption from astrosilicate materials are available with $R= 14000$ \citep{Rogantini2018}, important when $N_\mathrm{H} \geq 10^{23}\,\mathrm{cm}^{-2}$.

\begin{figure}[htp]
\centering
\includegraphics[width=0.49\textwidth]{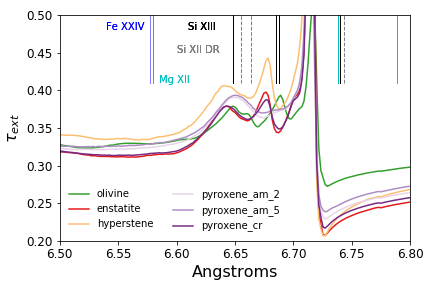}
\includegraphics[width=0.49\textwidth]{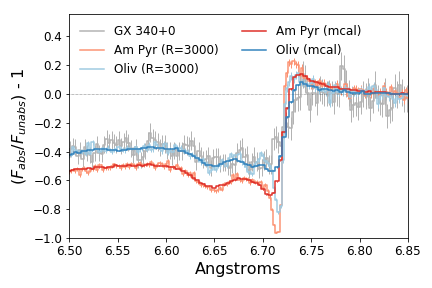}
\caption{
{\sl Left:} Silicate cross-sections for various mineral compositions, using optical constants from \citet{Zeegers2017}, publicly available with 1~eV resolution.
{\sl Right:} Simulated spectra for GX~340+0, a highly absorbed LMXB with strong Si K shell features. XAFS from a crystalline olivine (blue) and amorphous pyroxene (red), with microcalorimeter ($R \sim 1000$) and $R=3000$ resolution, are overlaid on the {\sl Chandra} spectrum (grey). An exposure time of 50~ks with $A_{e} = 1000$~cm$^2$ was used.
}
\label{fig:SiK}
\end{figure}

\vspace{-2.5ex}
\subsection{What is the composition and crystalline fraction of silicates? (1.8~keV)}
\label{sec:Silicate}
\vspace{-1ex}

Silicate dust produced by some red giants contain signs of crystallinity \citep{2002A&A...382..222M, 2011ApJ...733...93G}, yet infrared Si-O spectra from the diffuse ISM primarily indicate amorphous silicon \citep{2001ApJ...550L.213L, 2004ApJ...609..826K}, providing strong evidence for significant dust processing in the ISM.

Lab measurements of common silicate materials have been performed at $R \sim4000-7000$ \citep{Zeegers2017}. They show multiple absorption resonances associated with the crystalline spacing of Si (Figure~\ref{fig:SiK}, left). 
Some of these features are spread out or lost when the crystalline structure is destroyed, as seen in the features for amorphous pyroxene. The depth of the near-edge absorption peak also varies depending on the mineral. 
Using a cross-section that includes the effects of dust scattering, \citet{Zeegers2017} showed that the spectrum of low mass X-ray binary GX~5-1 contained features consistent with significantly more crystalline olivine than amorphous pyroxene, with grains as large as 0.5\,$\mu$m. 
This signature of crystalline olivine contradicts the IR spectroscopy paradigm of amorphous silicates dominating the diffuse ISM. Figure~\ref{fig:SiK} (right) shows how high throughput, high resolution X-ray spectroscopy can better differentiate between the two.

\vspace{-2.5ex}
\subsection{Mysteries of carbon dust growth and processing (0.3~keV)}
\label{sec:Carbon}
\vspace{-1ex}

Little work has been done on neutral C K shell absorption due to the fact that currently operating X-ray telescopes have an $A_{e}$ that drops quickly for $E < 0.5$~keV. 
This is unfortunate, as understanding the abundance and form of carbonaceous dust is currently one of the most active fields in astromineralogy. 
Small graphite and PAHs alone are considered responsible for the 2175~\AA\ absorption bump and emission features ranging from $3.4-20~\mu$m \citep{DL1984, DL2007}. 
However, models that utilize compound mixtures of silicates, amorphous carbon, and ices can also describe extinction and infrared emission features from the UV to sub-mm \citep{Zubko2004, Jones2013}.
Additionally, nanodiamonds are abundant in presolar grains found in meteorites, and IR spectra of circumstellar material around some Herbig Ae/Be stars also show signatures of nanodiamonds \citep[e.g.,][]{Jones2004,Goto2009}.

These hypotheses can be tested by X-ray observatories with  $A_{e} \geq 500$~cm$^2$ and $R \geq 4000$, as demonstrated for the case of methane, benzene, and nanodiamond material \citep{Bilalbegovic2018}. XAFS can also probe the process of hydrogenation of large organic molecules \citep{Reitsma2014}. Curently, these types of experiments are only feasible with X-ray gratings. 

\vspace{-2.5ex}
\subsection{What is the true abundance of metals in the local ISM?}
\label{sec:Absolute}
\vspace{-1ex}

Gas can be distinguished from solid phases only with high resolution X-ray spectra, providing a check on absolute abundances in the ISM -- a crucial assumption in measuring depletion. Furthermore, ISM absorption in the 0.3--10\,keV band is strongly influenced by the adopted mix of interstellar metals, affecting X-ray continuum models on the 30\% level \citep{Wilms2000}. 
A parametric study of silicate K~shell absorption in the spectra of low mass X-ray binaries demonstrated that the absorption edges are deeper than predicted by current X-ray ISM models, suggesting that Si abundances need to be increased \citep{Schulz2016}.
Balancing the dust and gas budget for ISM metals is fundamental for testing our understanding of cosmic star formation history and finding missing baryons, 
a key science charge for current and future X-ray observatories \citep{Bregman2007}. 

\vspace{-2.5ex}
\section{The current state and future prospects}
\label{sec:prospects}
\vspace{-1ex}

Only two X-ray observatories are currently capable of high resolution X-ray spectroscopy: \textsl{Chandra}, which excels in spectroscopic resolution but whose soft energy effective area has diminished significantly over time, and \textsl{XMM-Newton}, which does not have the necessary spectral resolution to distinguish solidly between grain types. 
Further, they do not have sufficient $A_{e}$ to allow selection of sightlines based on dust characteristics; rather, typically only the brightest, most absorbed X-ray sources are selected for dust studies. 
The next generation of X-ray telescopes are capable of $A_{e} \sim 500$--$2000\,\mathrm{cm}^2$, offering a factor of 10--100 improvement in signal-to-noise and opening a larger range of sight lines for scientific discovery. 

Planned X-ray observatories like the \textsl{X-ray Imaging and Spectroscopy Mission} (XRISM) and \textsl{Athena} will employ microcalorimeters to measure an X-ray spectrum from every pixel with 2--4\,eV resolution. 
Such instruments are ideal for measuring spectra from dust scattering halos, which will also exhibit fine structure that can 
be used to evaluate dust mineralogy and grain sizes. 
This effect is described in more detail in a separate white paper by Valencic et al.~(2019).

Unfortunately, at the soft end of the spectrum ($< 2$\,keV), where the majority of XAFS from interstellar metals appear, microcalorimeters can only achieve $R \sim 250$ at the O K edge ($R \sim 900$ for Si K). In addition, their effective areas diminish rapidly in the 0.3--0.5\,keV range where O and C absorption signatures arise. 
Recently developed critical-angle transmission (CAT) gratings technology can achieve $R \sim 2500-10^4$ in the 0.2--1.4\,keV bandpass \citep{CATgratings}. Such instruments are ideal for astromineralogy in our own Galaxy, but would also enable us to probe cold-phase oxygen in quasar absorption line systems out to $z = 1$ (carbon out to $z = 0.2$). 
\textbf{While X-ray microcalorimeters will do their part to reveal the mineralogy of silicate dust, the state of the two most abundant yet mysterious interstellar metals -- carbon and oxygen -- can more readily be studied with the latest advances in X-ray gratings spectroscopy. Furthermore, high resolution lab measurements of X-ray absorption from fundamental mineralogical building blocks, such as ices and organic molecules, are necessary to interpret the observations that will be made in the next 20 years.}

\pagebreak

\bibliographystyle{aasjournal}
\renewcommand{\bibsep}{0pt}
\renewcommand{\bibsection}{}
\noindent\textbf{References}
\begin{multicols}{2}
\bibliography{references_new,jwadd,dust_wp_add}
\end{multicols}
\end{document}